\documentclass[fleqn,10pt]{wlscirep}
\usepackage[utf8]{inputenc}
\usepackage[T1]{fontenc}
\usepackage{lineno}
\usepackage{multirow}
\usepackage{amsmath}
\usepackage{siunitx}
\usepackage{tikz}
\usepackage{makecell}
\usepackage{xurl}
\title{MSPB: a longitudinal multi-sensor dataset with phenotypic trait measurements from honey bees (\textit{Apis mellifera} L.)} 

\author[1,*]{Yi Zhu}
\author[1,]{Mahsa Abdollahi}
\author[2]{Ségolène Maucourt}
\author[3]{Nico Coallier}
\author[1]{Heitor R. Guimarães}
\author[2]{Pierre Giovenazzo}
\author[1]{Tiago H. Falk}
\affil[1]{INRS-EMT, Université du Québec, Montréal, Canada}
\affil[2]{Département de biologie, Université Laval, Laval, Canada}
\affil[3]{Nectar Technologies Inc., Montréal, Canada}

\affil[*]{corresponding author(s): Yi Zhu (Yi.Zhu@inrs.ca)}

\begin{abstract}
We present a longitudinal multi-sensor dataset collected from honey bee colonies (\textit{Apis mellifera}) with rich phenotypic measurements. Data were continuously collected between May-2020 and April-2021 from 53 hives located at two apiaries in Québec, Canada. The sensor data included audio features, temperature, and relative humidity. The phenotypic measurements contained beehive population, number of brood cells (eggs, larva and pupa), \textit{Varroa destructor} infestation levels, defensive and hygienic behaviors, honey yield, and winter mortality. Our study is amongst the first to provide a wide variety of phenotypic trait measurements annotated by apicultural science experts, which facilitate a broader scope of analysis. We first summarize the data collection procedure, sensor data pre-processing steps, and data composition. We then provide an overview of the phenotypic data distribution as well as a visualization of the sensor data patterns. Lastly, we showcase several hive monitoring applications based on sensor data analysis and machine learning, such as winter mortality prediction, hive population estimation, and the presence of an active and laying queen.
\end{abstract}

\begin{document}

\flushbottom
\maketitle

\thispagestyle{empty}


\section*{Introduction}
{\it{Apis mellifera}}, commonly referred to as honey bees, hold significant commercial value not only for the hive products but also due to their fundamental role as pollinators. Their pollination activities are beneficial for both agricultural crops and biodiversity~\cite{caas2020good, klein2007importance, aizen2009global}. Historically, beekeepers have relied on manual and visual inspections to monitor their beehives~\cite{abdollahi2022automated}, but these are time-consuming and disruptive to the colonies. While beekeepers typically inspect their hives on a regular basis (e.g., weekly to monthly) during pollination or honey production, important changes in colonies can occur within that time frame, making continuous monitoring essential~\cite{abdollahi2022automated}. Meanwhile, large-scale colony losses have been observed worldwide in recent years, caused by multiple stressors acting independently or synergistically, such as pesticides, pathogens, parasites, climate changes, and many other factors~\cite{gaubert2023individual, brodschneider2018multi, gray2019loss, gray2020honey}. Human inspections, however, cannot provide timely management for these conditions, and may lead to a greater amount of hive losses~\cite{capa}. 

Recently, computer-aided automated beehive monitoring systems have been developed to address the limitations of human management~\cite{abdollahi2022automated}. Existing systems typically place sensors inside the hive to record the environmental changes as well as colony status. High-level features are then extracted from sensor data then fed into machine learning (ML) models for downstream tasks, such as the early prediction of colony winter survivability~\cite{zhu2023bee}, estimation of colony strength~\cite{zhang2021semi}, and discrimination of different bee activities~\cite{terenzi2021comparison}, just to name a few. Among different sensor modalities, temperature and relative humidity are known as the two most widely used ones that are closely related to hive status. For example, studies have shown an increased amount of honey production and lower mortality rates under well-controlled temperature levels~\cite{cetin2004effects, southwick1982metabolic, ecology1985study}. The optimum relative humidity of a beehive varies between 50\% to 60\%, while higher or lower levels are shown to have an impact on brood development and mite infestation levels~\cite{human2006honeybees, oertel1949relative}. Abrupt changes in internal hive temperature and relative humidity have also been used to detect the presence of an active queen bee, as well as to predict swarming~\cite{ferrari2008monitoring, abou2017review}.

More recent studies have explored the use of acoustic sensors to infer the present state of the colony~\cite{abdollahi2022automated}. Honey bees contract their thoracic wing muscles creating a vibration that generates complex acoustic signals~\cite{kirchner1993acoustical, pastor2005brief}. Compared to conventional modalities, audio signals provide a more direct measurement of the hive status and reflects the instant response of bees to outer changes. Given the advantages of beehive acoustics, a substantial body of work has applied ML algorithms to audio data to detect the presence of a queen bee~\cite{uthoff2023acoustic, nolasco2019audio, kim2021acoustic}, swarming~\cite{uthoff2023acoustic, zgank2019bee, zgank2021iot}, as well as other activities~\cite{terenzi2019features, cecchi2018preliminary}. 

While automated beehive monitoring systems are advantageous in multiple aspects, massive amounts of data are needed to enable accurate ML model training and decision-making~\cite{jordan2015machine}. To this end, we curated the {\bf M}ulti-modal {\bf S}ensor dataset with {\bf P}henotypic trait measurements from honey {\bf B}ees (MSPB), which is composed of audio, temperature, and relative humidity data recorded from a large number of hives located in Québec, Canada during a one-year period. This paper presents a detailed description of the data collection procedure, sensor data pre-processing, data records, and our preliminary findings based on statistical analysis and ML-driven hive monitoring tools.

To highlight the novelty of our database, a comparison with existing publicly available databases is summarized in Table~\ref{tab:data}. It should be noted that with NU-Hive and OSBH, only subsets were found available at \url{https://zenodo.org/records/1321278} while the full versions are not publicly available. Therefore, the descriptions in Table~\ref{tab:data} correspond only to these subsets. Despite the multiple efforts made, it can be seen from Table~\ref{tab:data} that existing datasets are limited in terms of hive sample size, time range, number of sensor modalities, and variety of phenotypic trait measurements, thus making the development of ML tools challenging. As such, we introduce the MSPB dataset to tackle these limitations and to provide the research community with a richer dataset to help advance beehive monitoring. 

The MSPB database was collected non-stop for one year from May, 2020 to April, 2021, using 53 hives located at two apiaries in Québec, Canada. The audio, temperature, and relative humidity data were recorded synchronously throughout the whole year, resulting in a total of 365 days of data. Compared to existing publicly available databases, the MSPB dataset covers the longest time range and provides much richer phenotypic traits annotated by apicultural science experts, including the colony honey bee population, honey yield, queen-related conditions (e.g., swarming, supersedure, egg laying), health status (e.g., \textit{Varroa} mite infestation, winter survivability), and multiple behavioral evaluation results. These phenotypic trait measurements, together with the large number of hives, would allow for a more systematic analysis of sensor data to understand honey bee activities.

\begin{table}[]
    \centering
    \begin{tabular}{|c|c|c|c|c|c|c|}
    \hline
    \bf Dataset & \bf \#Hives & \bf Modality & \bf Location & \bf \#Days & \bf Annotations & \bf Availability\\
    \hline
    NU-Hive$^{\dagger}$~\cite{cecchi2018preliminary} & 2 & \begin{tabular}{@{}c@{}}Audio \\ Temp \\ Humid \\ CO\textsubscript{2} \end{tabular}  & IT & 5 & Queen presence & Subset\\
    \hline
    BUZZ$^{\ddagger}$~\cite{kulyukin2021audio} & 6 & Audio & UT, USA & 109 & Sound type (e.g., bee, noise, etc.) & Upon requests\\
    \hline
    OSBH$^{\dagger}$~\cite{nolasco2018bee} & 6 & Audio & EU; USA; AU & 2 & Queen presence & Subset\\
    \hline
    MSPB (Ours) & 53 & \begin{tabular}{@{}c@{}}Audio \\ Temp \\ Humid \end{tabular} & QC, CA & 365 & 
    \begin{tabular}{@{}c@{}}Population\\ Honey yield \\ Queen conditions \\ Defensive behavior \\ Hygienic behavior\\ Winter mortality
    \end{tabular} & Fully available \\
    \hline
    \end{tabular}
    \caption{Comparison of MSPB with other publicly available beehive sensor datasets. $^{\dagger}$NU-Hive and OSBH described here are publicly available subsets of their parental databases. $^{\ddagger}$BUZZ comprises four subsets (where BUZZ 3 is a subset of BUZZ 4), we report the metadata after aggregating all subsets.}
    \label{tab:data}
\end{table}

\section*{Methods}
\subsection*{Honey bee colonies}
\begin{figure}
    \centering
    \includegraphics[width=\linewidth]{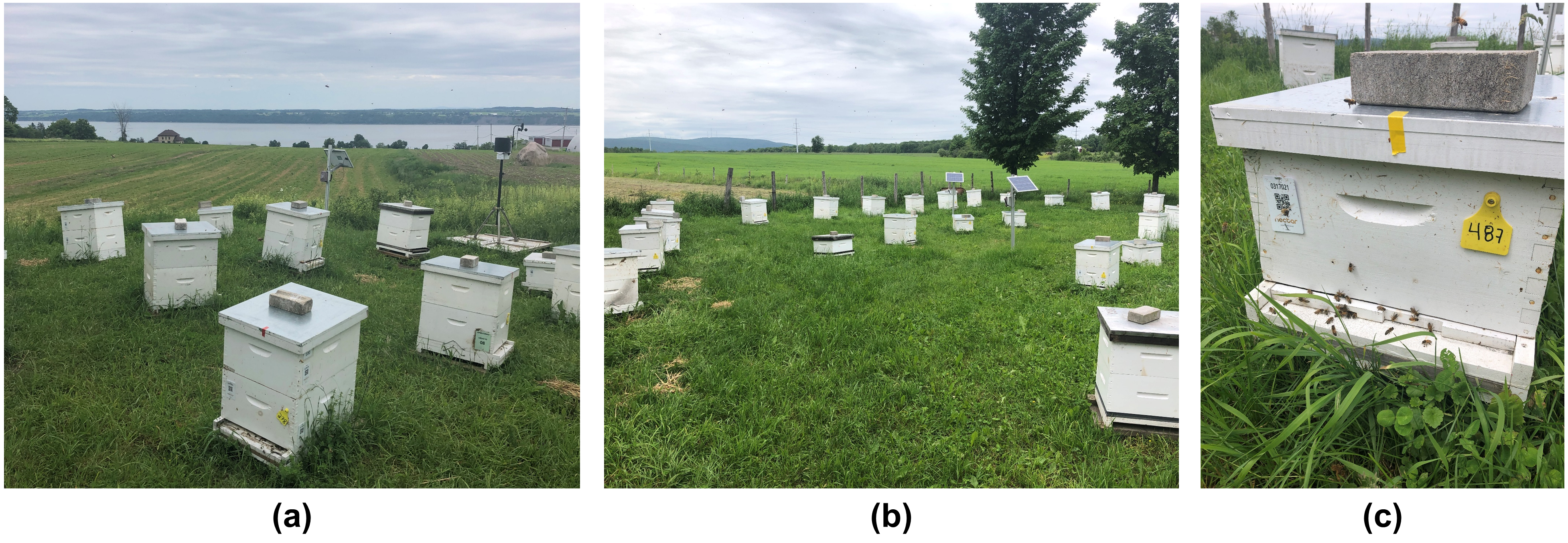}
    \caption{Photos of (a) Dubuc hives, (b) Côté hives, and (c) a closer view of one hive chamber.}
    \label{fig:ap}
\end{figure}

Our study was conducted with 53 honey bee colonies selected amongst the livestock of the Centre de recherche en sciences animales de Deschambault (CRSAD) Québec, Canada (N46$^{\circ}$40.270, W10$^{\circ}$71.500). Selected colonies had sister queens, were of equivalent strength (6–7 frames of bees/brood) and housed in 10-frame Langstroth hives mounted with a Plastic Varroa Stainless Steel Screen Bottom Board (Propolis-etc..., Saint-Pie, QC, Canada; SD-1500). Colonies were managed for honey production with a single brood chamber and placed in two farmland sites, i.e., the Dubuc (N46$^{\circ}$42.27, W71$^{\circ}$34.33) and Côté (N46$^{\circ}$44.302, W71$^{\circ}$28.284) apiaries. The Dubuc apiary is on a small hill and is exposed to more wind compared to the Côté Apiary. Figure~\ref{fig:ap} shows the photos of hives taken at the two apiaries, as well as a closer view of one hive with one brood chamber at the start of experiment.

\subsection*{Hive management}
\begin{figure}
    \centering
    \includegraphics[width=0.6\linewidth]{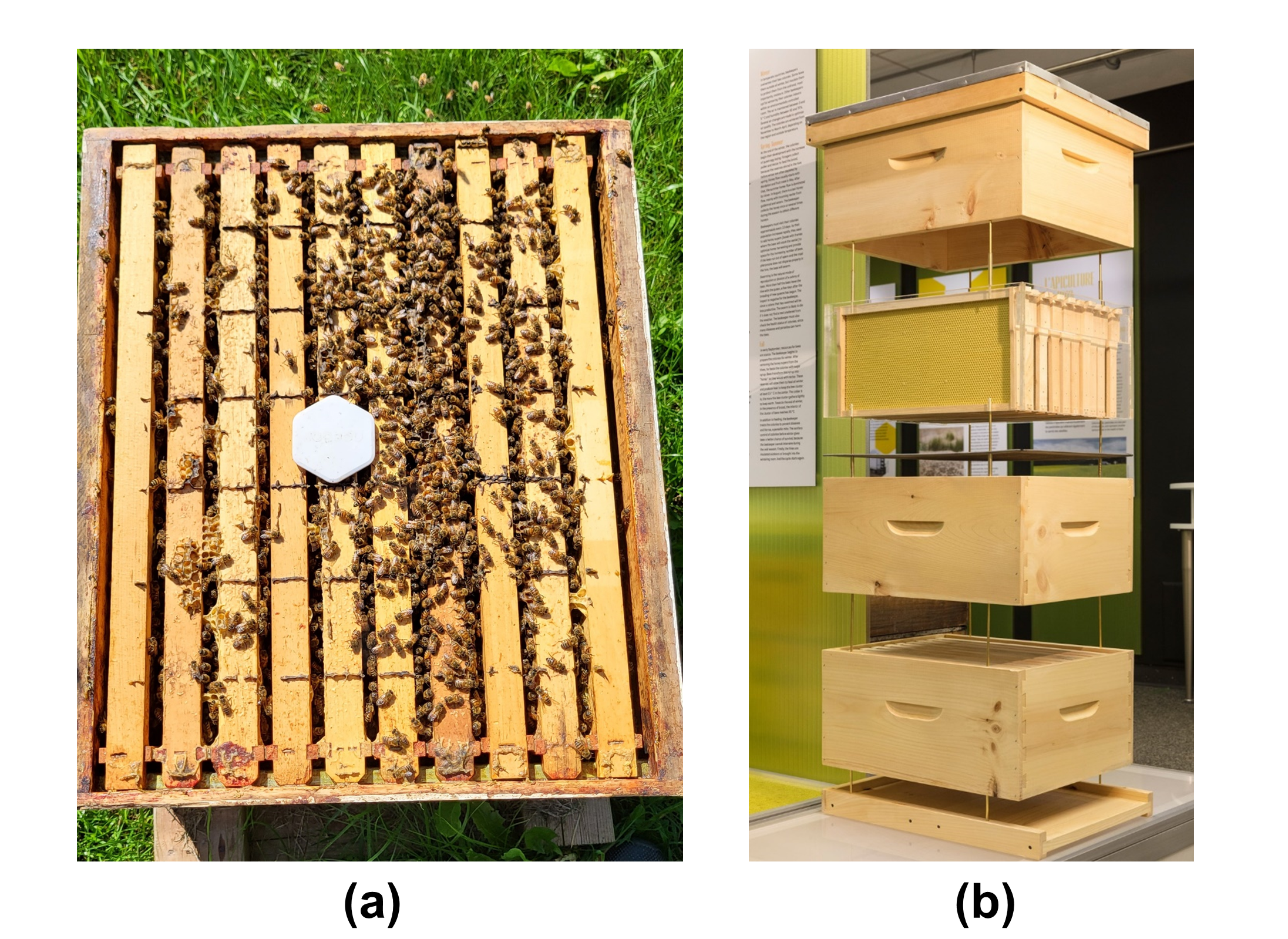}
    \caption{Illustrations of (a) the position of the sensor over the frames of the bottom brood chamber, and (b) the decomposed overall structure. From bottom-up: base board, two brood chambers, queen excluder, honey super showing frames and box, top cover roof.}
    \label{fig:chamber}
\end{figure}

During the summer, extra honey supers were added over the brood chamber and separated by a queen excluder. The positioning of the sensor on the top of the frames of the brood chamber and the overall structure of a hive are illustrated in Figure~\ref{fig:chamber} (a) and (b). At the beginning of September, honey supers were removed, and colonies had one brood chamber. Fall feeding started on  September 15, 2020 and all colonies were given 24 liters of a sucrose 2:1 solution using a top box feeder (Wooden Miller feeder \# FE-1100 at Propolis-etc, Beloeil, Québec). Colonies received a Thymovar anti-\textit{Varroa} treatment (Propolis-etc..., Saint-Pie, QC, Canada; TH-1110), applied as per label, starting on September 17, 2020, followed by an oxalic acid treatment (Propolis-etc..., Saint-Pie, QC, Canada; AO-1201) on October 28, 2020 (drip method: \SI{35}{g.L^{-1}} in a sucrose 1:1 solution, \SI{5}{ml} between frames of the hive body crowded with honey bees). Colonies were wintered indoors in an environmentally controlled room (4-5C, 50-60 \% relative humidity) from November 14, 2020 to April 19, 2021 and then moved into a spring apiary in Deschambault, Québec near the bee research facility until mid-May.

\subsection*{Beehive monitoring system overview}
\begin{figure}
    \centering
    \includegraphics[width=0.65\linewidth]{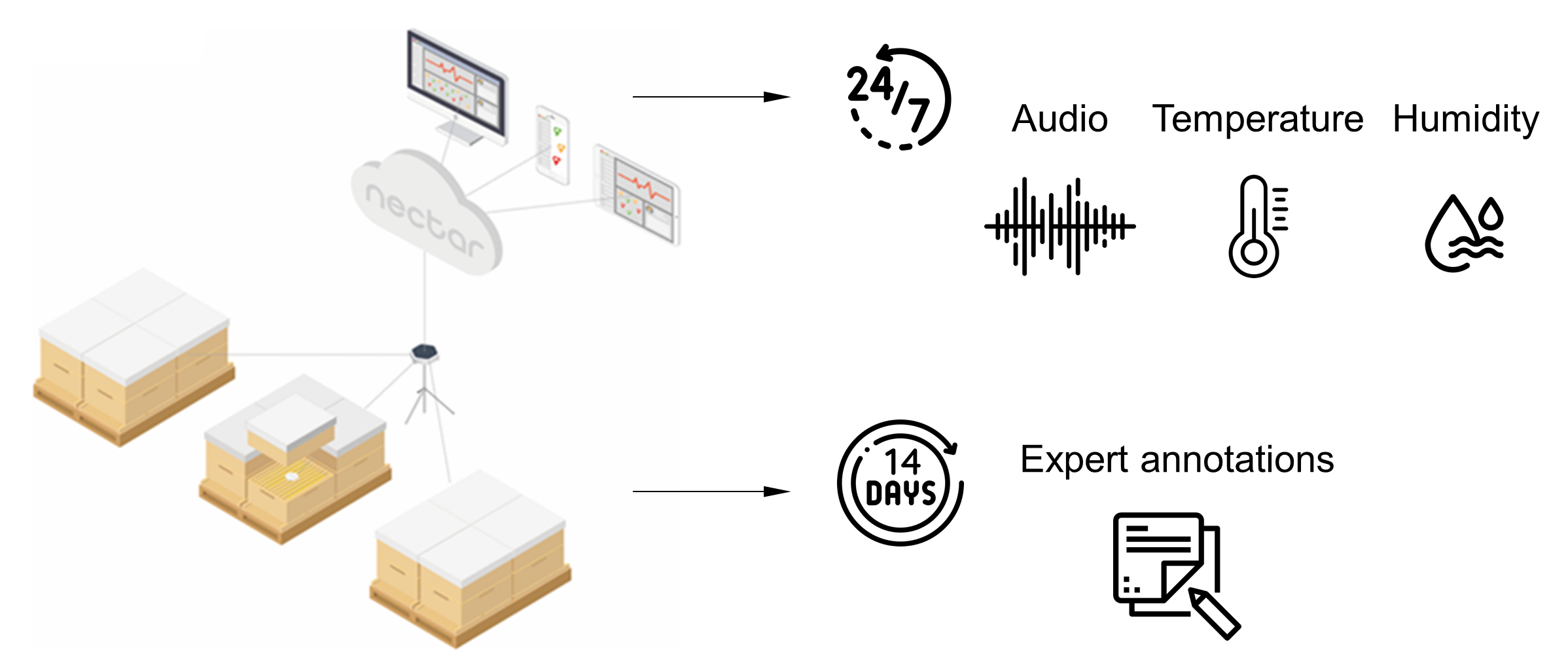}
    \caption{An overview of the beehive monitoring system with continuous multi-modal data recording and phenotypic traits. Sensor data were collected 24/7 at a fixed interval of \SI{5}{min} for audio, and \SI{15}{min} for temperature and relative humidity. Phenotypic traits were annotated by apicultural science experts every two weeks.}
    \label{fig:system}
\end{figure}

The beehive monitoring system comprises two fundamental components: (1) a multi-modal sensor system with continuous data recording, and (2) phenotypic traits annotated by apicultural science experts on a bi-weekly basis. An overview of the beehive monitoring system can be seen in Figure~\ref{fig:system}. With the sensor data collection, a multi-modal sensor is positioned at the top of the central frame of the brood box of a Langstroth hive housing honey bees (see Figure.~\ref{fig:chamber}a). This sensor is capable of concurrent recording of audio, relative humidity, and temperature data at regular intervals of \SI{5}{min}, \SI{15}{min}, and \SI{15}{min} respectively. The recorded sensor data is wirelessly transmitted to a central data aggregator powered by solar energy and securely stored in the cloud. The sensor data was collected 24 hours a day, 7 days per week from May 2020 to June 2021. Besides continuous sensor recording, apicultural science experts visited hives bi-weekly to monitor the hive status and conducted evaluations on a regular basis. Colony phenotypic trait measurements, such as honey bee population, honey yield, and health status, were also collected, hence providing valuable context to interpret the sensor data.

\subsection*{Phenotypic trait measurements}
Five phenotypic trait measurements were collected and are summarized in Table~\ref{tab:phenodata}. Details about the collection procedure are outlined as follows:

\begin{table}[]
    \centering
    \begin{tabular}{|c|c|c|}
    \hline
    {\bf Data type} & {\bf Collection date} & {\bf Description} \\
    \hline
    Brood surface & 20-06-09 & Number of capped, uncapped, and total cells in the 2 brood chambers\\
    \hline
    Honey production & 20-11-06 & Weight of honey produced by each colony \\
    \hline
    Defensive behavior & 20-07-21 \& 20-08-04 & Number of stings on a leather flag waved over the chamber for 2 minutes \\
    \hline
    Hygienic behavior & 20-07-23 \& 20-08-06 & Liquid nitrogen freezed-brood test and evaluation after 24h\\
    \hline
    \textit{Varroa} infestation level & 20-05-06 \& 20-08-13 & Evaluation of the \textit{Varroa} infestation level with the \textit{Varroa} wash technique \\
    \hline
    \end{tabular}
    \caption{Types of phenotypic trait measurements annotated by apicultural science experts.}
    \label{tab:phenodata}
\end{table}

\begin{itemize}
    \item {\bf Spring colony development}. Colony strength is measured by the area (cm2) occupied by immature worker honey bees (eggs + larvae + capped brood) in May-June. \cite{giovenazzo2011evaluation, delaplane2013standard}.
    \item {\bf Honey yield}. Colonies are equipped with honey supers, placed above a queen excluder. Each colony has at least two honey supers during honey flow, each with stretched comb frames.  Each colony was weighed before and after harvest using a platform scale (CAS-USA, East-Rutherford, NY, USA; CAS CI-2001BS), where the weight difference was considered as the honey yield.
    \item {\bf Hygienic behavior}. Hygienic behavior is evaluated with a test that measures the cleaning capacity of a colony's bees on a percentage level. During the test, the colony was opened, and a comb was selected containing a solid slab of sealed worker brood in the pupal stage, with pupae having pink or purple eyes. Two PVC tubes (\SI{5.08}{\centi\metre} internal diameter) were pressed to the comb's midrib. The number of empty (i.e., missed) cells in each tube was counted. Liquid nitrogen was then applied at a rate of \SI{300}{ml} per tube to freeze the brood. Frames were marked and returned to the colony. After 24 hours, the number of cells removed was counted and divided by the total number of cells, the resultant percentage value was then used as a measure of the hygienic behavior \cite{spivak1998field, spivak1998honey}. This test was carried out twice during the 2020 summer, at the honey flow low period, i.e., during a period when nectar resources are poor for bees, early August. 
    \item {\bf Defensive behavior}. The flag stinging test was used to measure a colony's defensive behavior. A flag was waved rhythmically (amplitude approximately \SI{20}{\centi\metre}), with an oscillation every \SI{2}{s}, \SI{5}{} to \SI{10}{\centi\metre} above the colony's brood chamber for 2 minutes. The flag is a piece of black suede leather (measuring 10 x 8 cm) suspended from a piece of light wood (0.7 x 0.5 x 100 cm). After the test, the number of stings on the flag was counted \cite{guzman2003relative, guzman2007elemental}.
    \item  {\bf \textit{Varroa} infestation level}. Two methods were employed to evaluate the \textit{Varroa} infestation level at the end of August. The first was the natural mite-fall method, where the {\textit{V. destructor}} infestation level was assessed by the natural mite-fall method using sticky boards placed on the bottom boards of hives \cite{imdorf2003alternative, dietemann2013standard}. Mites that had fallen on sticky boards (5–7 days) were counted to obtain a daily mite drop value. The second followed the alcohol washing method. This method involved sampling around 200 to 300 bees from a colony honey frame in 70\% alcohol. For each sample, the bees were counted and then placed in a \textit{Varroa} EasyCheck or \textit{Varroa} Mite Test Bottle type sampling device and covered with 70\% ethanol. The sampling device was then placed on a horizontal shaker at 150 rpm for 5 minutes. The bees were then removed and the number of mites in the sampling device was counted. The process was repeated until there were no more mites in the sample (up to 3 washes). The total number of \textit{Varroa} mites counted gave the number of \textit{Varroa} mites per 100 bees (i.e., total number of \textit{Varroa} mites in the sample$\times$100/number of bees in the sample) \cite{de1982comparative, borba2022phenomic}.
\end{itemize}

\subsection*{Sensor data collection and pre-processing}
Table~\ref{tab:multisensor} presents the sensor data modalities, frequency in which parameters were extracted, and in the case of the audio modality, the microphone sampling frequency. To optimize bandwidth, battery life, and storage requirements, the audio data were not stored in the raw waveform format. Instead, four types of acoustic features were computed to encapsulate relevant information, including hive power, audio band density ratio, audio density variation, and audio band coefficients. Audio signals were originally sampled at \SI{15625}{\Hz}, the fast Fourier transform (FFT) was subsequently computed over 30 non-overlapping frames once every \SI{5}{\min}, each frame with a length of 512 points (i.e., \SI{0.98}{\s} of audio data considered every \SI{5}{\min}). We denote the resultant spectrogram as $X_{j,k}$, where $j$ corresponds to the frame index ($j \in \{0,1,...,29\}$), and the frequency bin index as $k$ ($k \in \{0,1,...,256\}$). 

The hive power ($P_{hive}$) reflects the overall power between \SI{122}{Hz} and \SI{515}{Hz}, where
\begin{gather}
    ABD_j  = \sum_{k=4}^{k=17}||{X_{j,k}}||^{2}  \ \text{and}\\
    P_{hive}  = 10\lg(\frac{\sum_{j=0}^{j=29}ABD_j}{30 \times (17-4+1)}) = 10\lg(\frac{\sum_{j=0}^{j=29}ABD_j}{420}).
\end{gather}
Here, $ABD_j$ corresponds to the audio band density at the time frame $j$, and the extra division by 14 in the denominator normalizes the energy for a per-bin representation, but can be removed by a single shift in dB. The selection of such frequency range was based on our exploratory analysis, where non-bee sounds generally manifest at different frequency ranges (e.g., rainfall) or exhibit abrupt fluctuations in signal power (e.g., human speech and thunder sounds). The audio band density ratio ($ABD_{ratio}$) is defined as the ratio of hive power with regard to the power of the whole frequency range ($AD$):
\begin{gather}
    AD_j = \sum_{k=4}^{k=256}||{X_{j,k}}||^{2} \\
    ABD_{ratio} = \frac{ABD_j}{AD_j}
\end{gather}
The audio band density variation ($ABDR$) reflects the amount of changes within the \SI{0.98}{s}:
\begin{gather}
    ABDR = 10\lg{\frac{\max({AD_j})}{\min({AD_j})}}
\end{gather}
For a consistent audio event like a bee sound, this value will be at a lower \SI{}{dB}, while for other events such as thunder or human speech, the density variation is expected to be at a higher value. Lastly, we chose 16 linearly spaced frequency bins and computed the power, respectively, as the 16 coefficients:
\begin{gather}
    Bin_{N} = \frac{N \times 15625}{512}, n \in \{4, 5, 6, ..., 19\}, \\
    Coef_{N} = 10\lg{(\frac{\sum_{j=0}^{29}||X_{j,N}||^2}{30})}.
\end{gather}
The aforementioned acoustic features were extracted on-the-fly and stored as the final format on the cloud server. As for relative humidity and temperature, the raw data were preserved in percentage and degrees, respectively, where higher values corresponding to increased relative humidity and higher temperatures within the beehive boxes.

\begin{table}[]
    \centering
    \begin{tabular}{|c|c|c|c|c|}
    \hline
    Modality & Freq. & Fs &  Feature & Description\\
    \hline
    \multirow{4}{*}{Audio} & \multirow{4}{*}{\SI{5}{min}}& \multirow{4}{*}{\SI{15625}{Hz}} & Hive power &  Overall power within [\SI{122}{Hz}-\SI{519}{Hz}] in \SI{}{dB}\\ \cline{4-5}
    & & & Audio Band Density Ratio & Ratio of hive power over the overall power \\ \cline{4-5}
    & & & Audio Density Variation & Amount of changes in audio power within \SI{0.98}{s}\\ \cline{4-5}
    & & & Audio Band Coefficients & Power of 16 linearly spaced frequency bands in \SI{}{dB}\\
    \hline
    Temperature & \SI{15}{min} & - & Raw value in degree & Temperature insides the chambers\\
    \hline
    Relative Humidity & \SI{15}{min} & - & Raw value in percentage & Relative humidity insides the chambers\\
    \hline
    \end{tabular}
    \caption{List of stored features computed from multi-modal sensor data. Temperature and relative humidity data were stored as raw values, while audio data were stored as acoustic features.}
    \label{tab:multisensor}
\end{table}

\section*{Data Records}


The MSPB dataset is made fully available at the Zenodo repository \url{https://doi.org/10.5281/zenodo.8371700}. The sensor data and phenotypic traits were stored separately in \url{.csv} format, each of which was further divided into two files based on the time range, resulting in a total of four \url{.csv} files. To distinguish summer and winter data, those collected between April, 2020 and October, 2020 received a `D1' label in the file name, while the data between October, 2020 and April, 2021 were labelled as `D2'. The detailed file composition is summarized in Table~\ref{tab:file-summary}. The total size of the shared files is about \SI{500}{\mega\byte}. 

D1 and D2 sensor data are both paired with (1) the time stamp (date and time) of the data collection, (2) hive ID, which is a unique number to identify each hive, (3) apiary ID, which indicates the apiary location of the hive, (4) temperature values, (5) relative humidity values, and (6) twenty audio features. The D1 phenotypic traits file has three sub-sheets, which details (1) the visit date and time of the human evaluations, as well as the evaluation tasks, (2) the population size of the colonies measured at each visit, (3) other phenotypic trait measurements, such as \textit{Varroa} infestation levels, defensive and hygienic behavior, honey yield, etc. During the period of D2, hives were maintained in the winter chambers and only evaluated once in the Spring to check their winter survival rate. Hence, the D2 phenotypic traits file contains the survival status, as well as the mortality causes (if any) of all hives.

\begin{table*}[]
\centering
\begin{tabular}{c|c|c|c}
\hline
File name & Sub-sheets & Variables & Description \\
\hline
\multirow{6}{*}{\makecell{\url{D1_sensor.csv} \& \\ \url{D2_sensor.csv}}} & \multirow{6}{*}{-} & Published\_at & Time stamp (\textsc{YYYY-MM-DD HH24:MI:SS})\\
& & Hive ID & ID unique to each hive\\
& & Apiary ID & ID of the apiary \\
& & Temperature & Temperature in degree Celsius\\
& & Relative humidity & Humidity in percentage \\
& & Audio features & twenty audio feature values \\
\hline
\multirow{21}{*}{\url{D1_ant.xlsx}} & \multirow{6}{*}{Visit} & Yard & Name of the apiary \\
& & Dates & Date of the visits \\
& & Arrival time & Time of arrival (\textsc{HH-MM})\\
& & Departure time & Time of departure (\textsc{HH-MM})\\
& & Manipulations & Details of human evaluations \\
& & Yard location & Coordinates of two apiaries \\
\cmidrule{2-4}
& \multirow{6}{*}{Evaluation [NUMBER]} & Dates & Date of the population evaluation\\
& & Yard & Name of the apiary \\
& & Hive ID & ID unique to each hive \\
& & Number of boxes & Bottom chamber + honey supers \\
& & NoF covered by bees & Population measurement \\
\cmidrule{2-4}
& \multirow{5}{*}{Phenotypic measurements} & Brood surface & Capped, uncapped, and total brood cells \\
& & \textit{Varroa} infestation & Severity of \textit{Varroa} \\
& & Defensive behavior & Defensiveness measured by flag test \\
& & Hygienic behavior & Cleaning capacity \\
& & Honey weight & Total honey produced during summer \\
\cmidrule{2-4}
& READ ME & - & A quick introduction on file content of all sub-sheets\\
\hline
\multirow{6}{*}{\url{D2_ant.xlsx}} & \multirow{6}{*}{-} & Apiary & Name of the apiary \\
& & Hive ID & ID unique to each hive\\
& & Mortality cause & Causes of the failed hives \\
& & Weight & Weight values (\SI{}{kg}) before and after winterization \\
& & Bees frames & Population before and after winterization \\
& & Syrup consumption & Syrup consumption (\SI{}{kg}) during winterization\\
\hline
\end{tabular}
\caption{Structure of the multi-modal sensor data and phenotypic trait measurement files.}
\label{tab:file-summary}
\end{table*}

\section*{Technical Validation}

\subsection*{Phenotypic trait representation}
\begin{figure*}
\centering
\includegraphics[width=\linewidth]{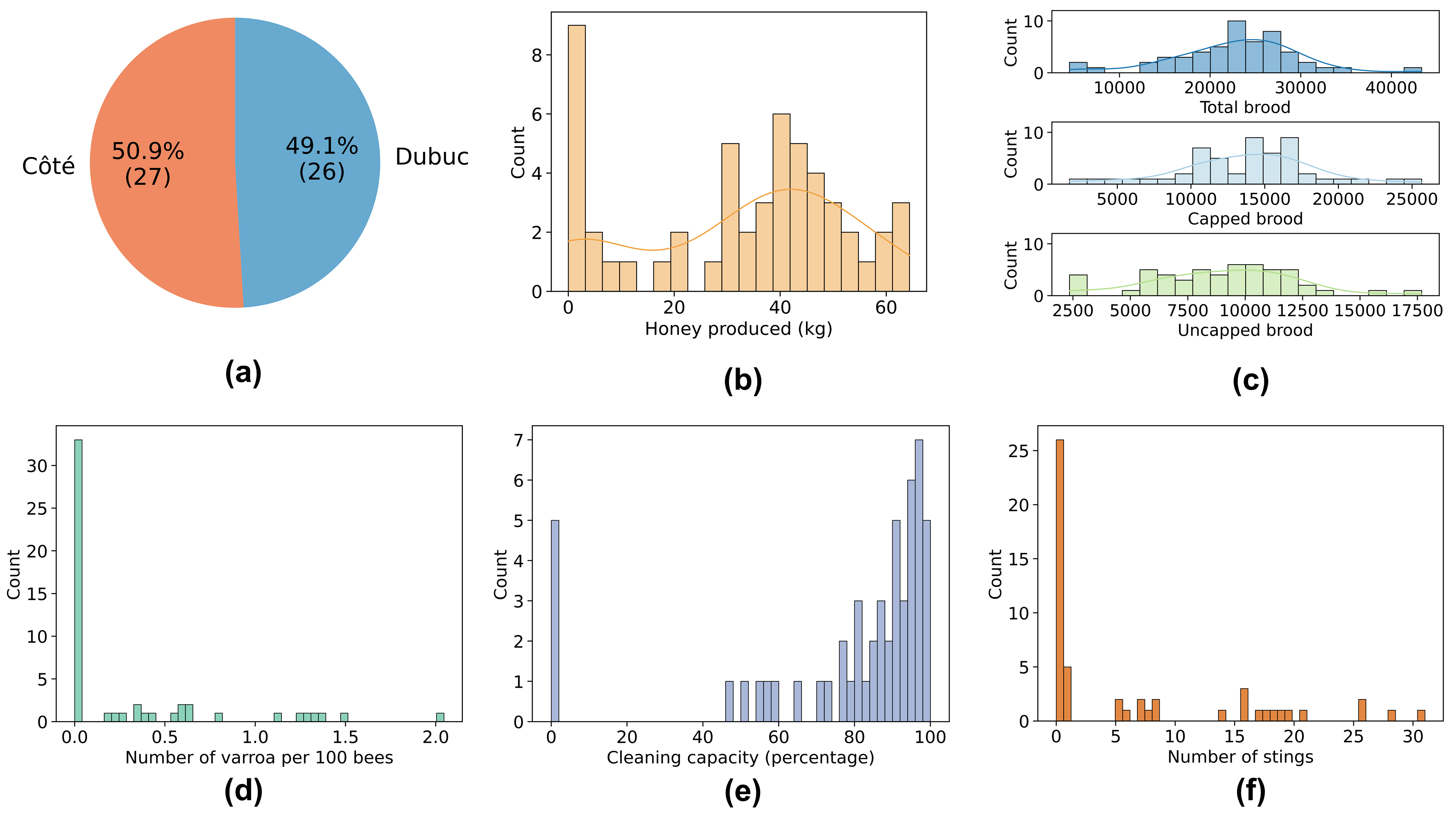}
\caption{Distribution of (a) number of hives across apiaries, (b) honey yield, (c) Brood population, eggs, larvae and pupa, during spring colony build up in June, (d) \textit{Varroa} infection levels, (e) Hygienic behavior, and (f) defensive behavior.}
\label{fig:pheno1}
\end{figure*}

The phenotypic traits of all bee hives from both apiaries are depicted in Figure~\ref{fig:pheno1}. Hives are equally distributed across two apiaries, each with 26-27 hives (see Figure~\ref{fig:pheno1}a). In terms of total honey production, the majority produced \SI{30}{} to \SI{60}{kg}, while several hives produced less than \SI{10}{kg} of honey. For most of the cases, the low productivity was related to queen bees. For example, there were five hives that went through queenless conditions, which led to the division or failure of the entire colony. Regarding bee population, the average of total brood (i.e., eggs, larva and pupa of honeybees) was approximately 25,000, with the majority varying between 20,000 to 40,000 (see Figure~\ref{fig:pheno1}c). During the summer, bee experts also evaluated the \textit{Varroa} condition, cleaning capacity, and hive defensive behavior on a regular basis. We calculated the average from all evaluations and summarized these data in Figure~\ref{fig:pheno1}d-f respectively. At the end of August \textit{Varroa} infestations were below the economic threshold level of 3

\begin{figure}
    \centering
    \includegraphics[width=0.5\linewidth]{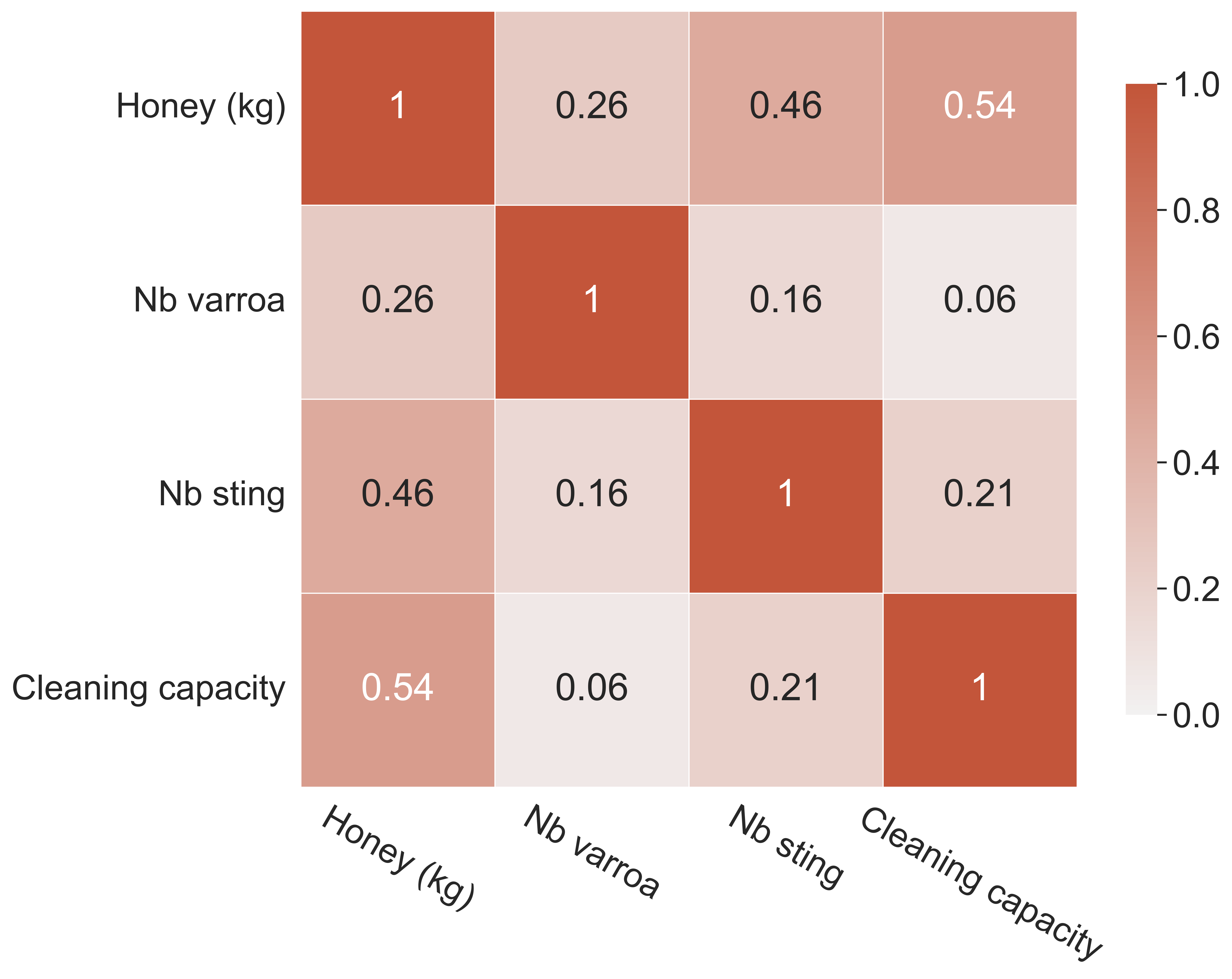}
    \caption{Pearson correlation coefficients between different phenotypic trait measurements. A strong statistical significance is found between honey yield and cleaning capacity, and defensive behavior (i.e., number of stings).}
    \label{fig:hm}
\end{figure}

To investigate the associations between different colony phenotypic traits, the Pearson correlation coefficient \textit{r} was calculated between honey yield, \textit{Varroa} infection, hygienic behavior, and defensive behavior (see Figure~\ref{fig:hm}). The honey yield was shown to be strongly associated with the hygienic behavior (\textit{r}(51) = .54, \textit{t} = 4.58, \textit{p} < .005) and defensive behavior (\textit{r}(51) = .46, \textit{t} = 3.70, \textit{p} < .005), while a weaker but significant correlation was found with \textit{Varroa} mite infection (\textit{r}(51) = .26, \textit{t} = 1.93, \textit{p} < .05). Meanwhile, no significance was found with the correlations between \textit{Varroa} infestation, cleaning capacity, and defensive behavior (\textit{r} $<0.3$). Together, our results suggest that the cleaning capacity and defensive behavior are the two factors that are most closely associated with honey yield.

\subsection*{Temporal patterns of multi-modal sensor data}
Similar to weather changes, the behaviors of bees follow a specific pattern that repeats on a daily and yearly level. Here, we show that such patterns can be captured by the multi-modal sensor data. With the yearly pattern, we first aggregated data points from each day and calculated the daily average for each hive. We then computed the mean and standard deviation across all hives to obtain the general pattern throughout the year. With the daily pattern, data points were aggregated hourly and averaged across the whole year. An average was then computed across all hives to obtain the \SI{24}{H} changes.
\subsubsection*{Yearly changes}
\begin{figure}
\centering
\includegraphics[width=\linewidth]{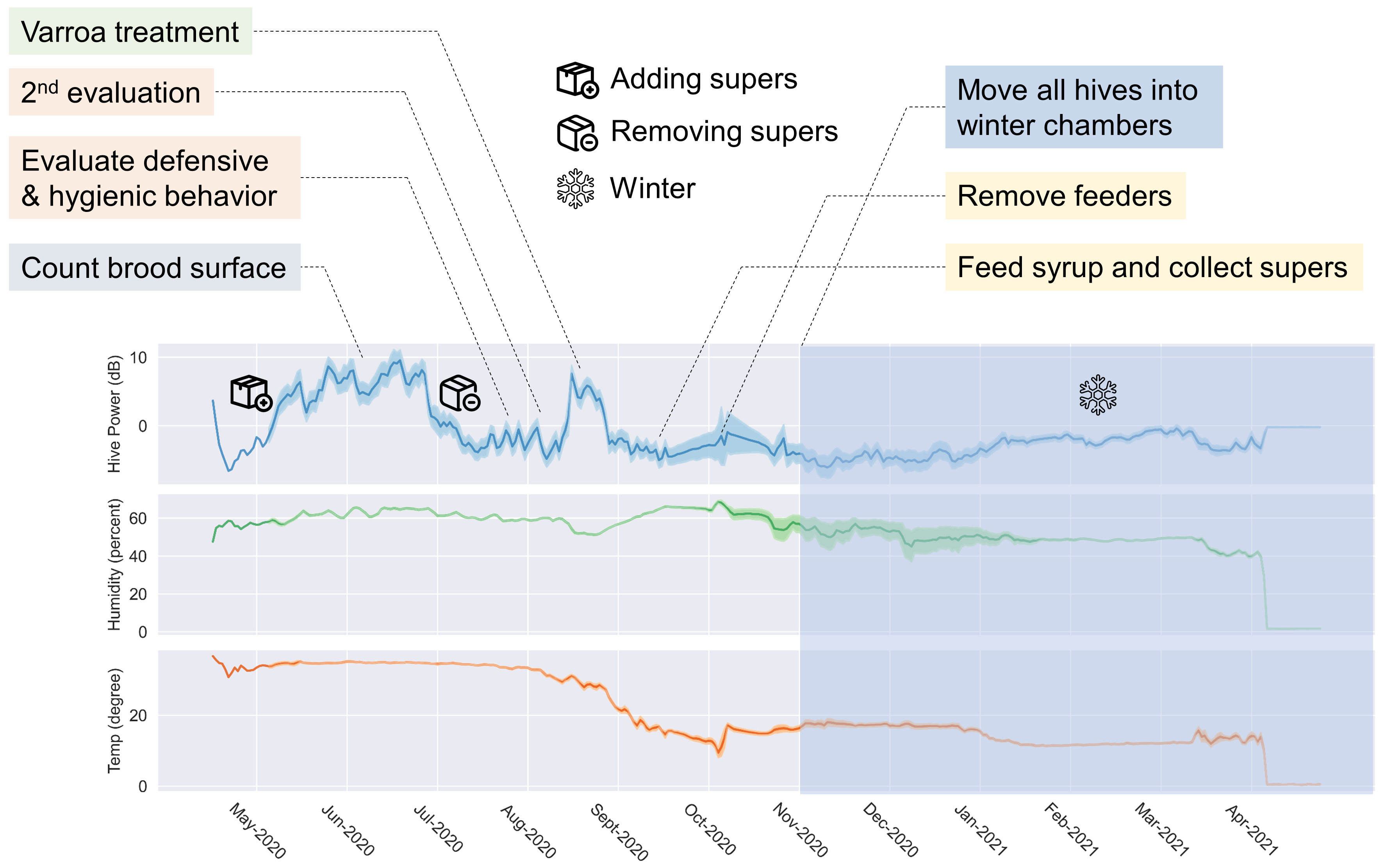}
\caption{Multi-sensor data from May-2020 to April-2021. Human evaluations are annotated on top of the plots. The lines are formed by connecting daily data points averaged across all hives. The shaded area represents one standard deviation from the average power of all hives.}
\label{fig:yearly}
\end{figure}

The changes of multi-modal sensor data from May 2020 to April 2020 are depicted in Figure~\ref{fig:yearly}, annotated with the honey bee experts evaluations and interventions. Compared to temperature and humidity, the audio modality exhibits greater variations across time, many of which are conditioned on human activities. For example, the hive power increases and decreases accordingly when the number of supers were being added or reduced to each hive. Peaks in hive power are also seen during evaluations and treatments, such as during the Thymovar\textsuperscript \textregistered \textit{Varroa} treatment in mid-September and two behavior evaluations before and after August 1st. After being moved to the winter chambers, all hives manifest a similarly steady pattern, which is reflected by the reduced variation across time and the smaller standard deviation across hives. Furthermore, hive power also exhibits higher standard deviations across hives compared to the other modalities, indicating that bee acoustics might encapsulate richer information unique to each hive. In general, the sensor data are found to be a good indicator of the arousal level of the bees, especially the audio.

\subsubsection*{Daily changes}
\begin{figure}
\centering
\includegraphics[width=0.85\linewidth]{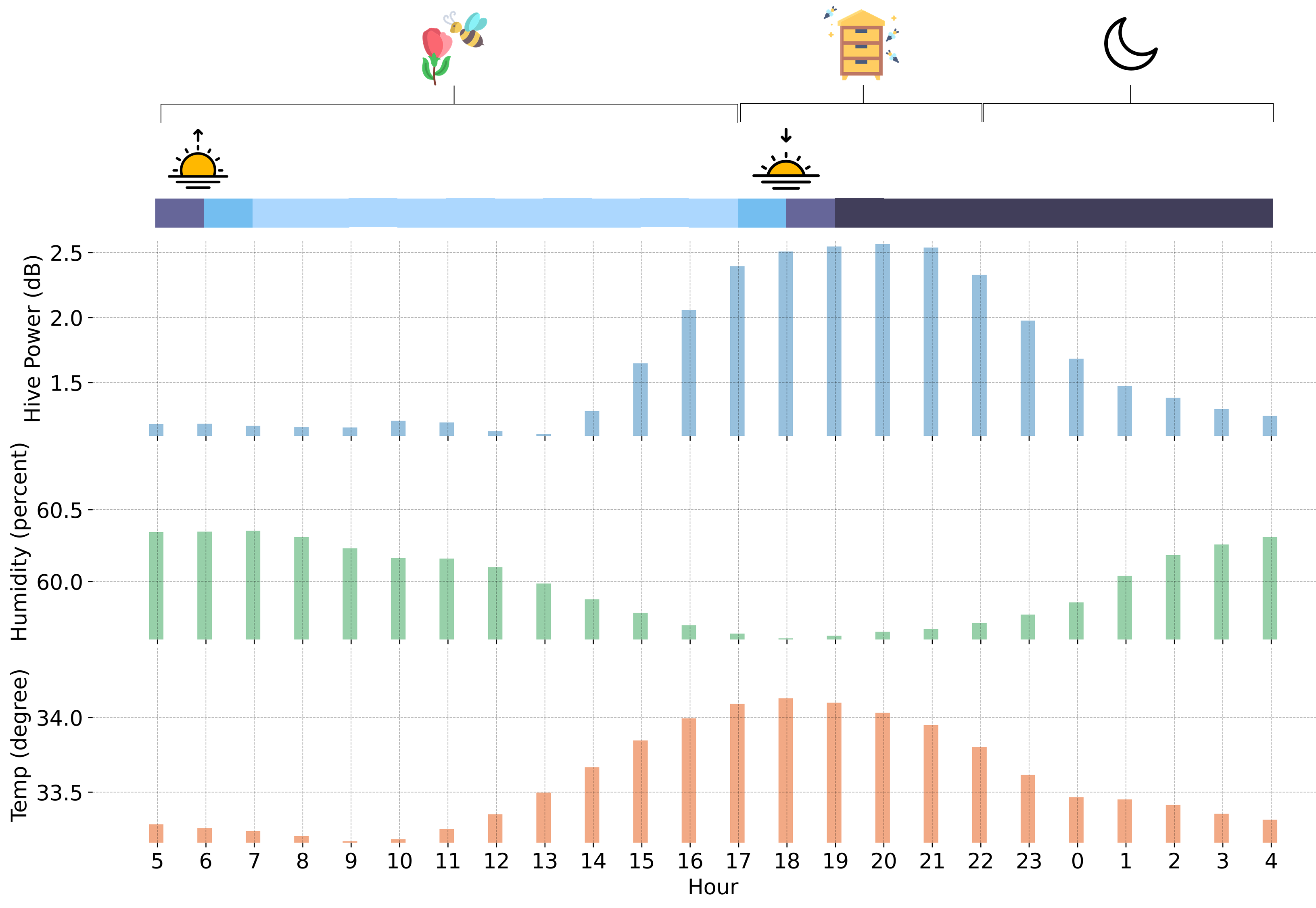}
\caption{24-hour changes of multi-sensor data together with the sunrise/sunset time. During the summer months, the sunrise time varies from 5AM to 7AM and the sunset time from 8PM to 6PM. The honeybee circadian clock is illustrated at the top of the plot, suggesting changes in bee behaviors across different times of the day.}
\label{fig:hourly}
\end{figure}

Honeybees are known to follow a particular daily routine driven by their circadian clock~\cite{moore2001honey}. Figure~\ref{fig:hourly} shows the summer-averaged \SI{24}{h} pattern in hive power, relative humidity, and temperature, together with the sunrise and sunset time across all hives. From sunrise, bees leave the hives to forage, resulting in a relatively lower hive power and relative humidity, as well as reduced temperature due to the absence of bees. After 3-4PM, a rapid increase is seen with hive power and temperature, suggesting the return of bees. The peak of hive power appears subsequently at 6-9PM, indicating that the majority of bees have returned and may have started cleaning honeycomb cells. During the resting time (after 9PM), the hive power and relative humidity gradually decreases while the temperature increases simultaneously. In summary, the sensor data are found to be highly correlated with the daily activities of honey bees, which can be used to infer and monitor their behaviors.

\subsection*{Identifying queen cell related events from audio}
Queen cell is a special type of cell with an elongated shape, where a larva develops and matures into a new queen \cite{scheiner2013standard}. There are usually two distinct reasons for the existence of queen cells, namely supersedure and swarming. Supersedure cells are made when the queen is old, ill, or missing, hence a new queen is needed to replace the old queen. Swarming cells, on the other hand, occur when the colony reproduces in summer when resources are abundant and the colony is fully developed. As a result of swarming, a portion of the colony will leave with the existing queen while the new queen will be raised and stay at the current hive. The identification of the presence of queen cells is crucial for beekeepers since timely management is needed for both swarming and supersedure conditions. 

\begin{figure}
    \centering
    \includegraphics[width=\linewidth]{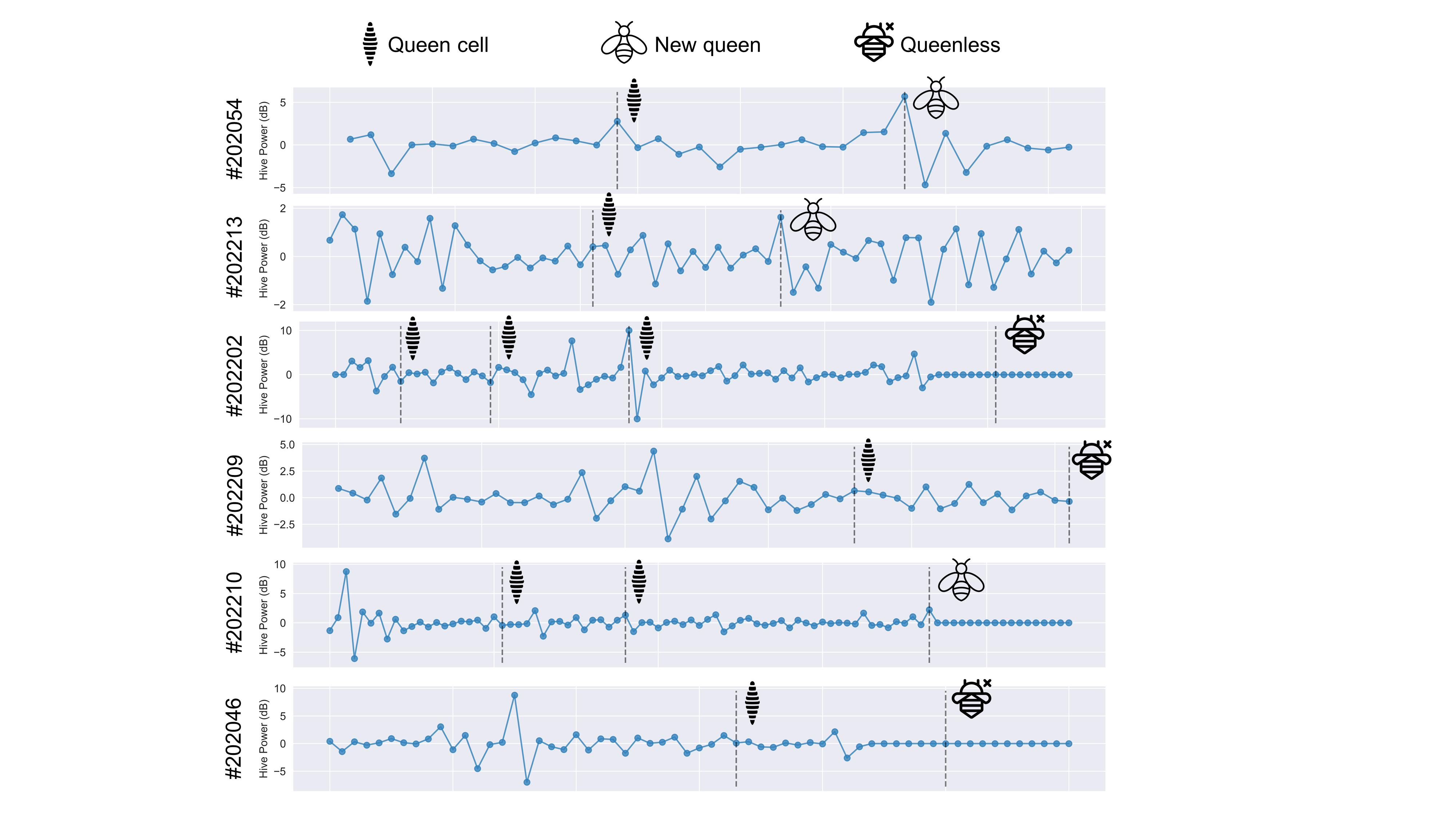}
    \caption{1st-order delta of the daily standard deviation of hive power, annotated with the swarm queen cells, new queen, or queenless conditions.}
    \label{fig:queen}
\end{figure}

Traditional methods rely on opening the hive and conducting visual observations (e.g., brood frames, population, etc.), which can be time-consuming and may lead to false conclusions or miss the best window of time for taking actions. Here, we demonstrate the potential of using sensor data to help identify the existence of queen cells. We hypothesize that the colony may manifest higher level of arousal close to the queen cell date, which can be reflected by an abnormal variation of hive power throughout the day. Among all 53 colonies, swarm queen cells were found in six of them. We calculated the standard deviation of hive power per hour and further computed the first-order differential to estimate the relative changes to the previous hour. The patterns of the acoustic descriptors of these six colonies are depicted in Figure~\ref{fig:queen}. With four colonies (\#202054, \#20223, \#202202, \#202210), peaks can be seen around the dates where the swarm queen cells, as well as the new queens, were found. However, the consistency of such pattern needs to be further investigated as peaks are not found in colonies \#202209 or \#202046. Notwithstanding, the analysis demonstrates that descriptors extracted from beehive acoustics can be a potential indicator of swarm queen cell related events.

\subsection*{Early detection of beehive winter survivability}
\begin{figure}
\centering
\includegraphics[width=\linewidth]{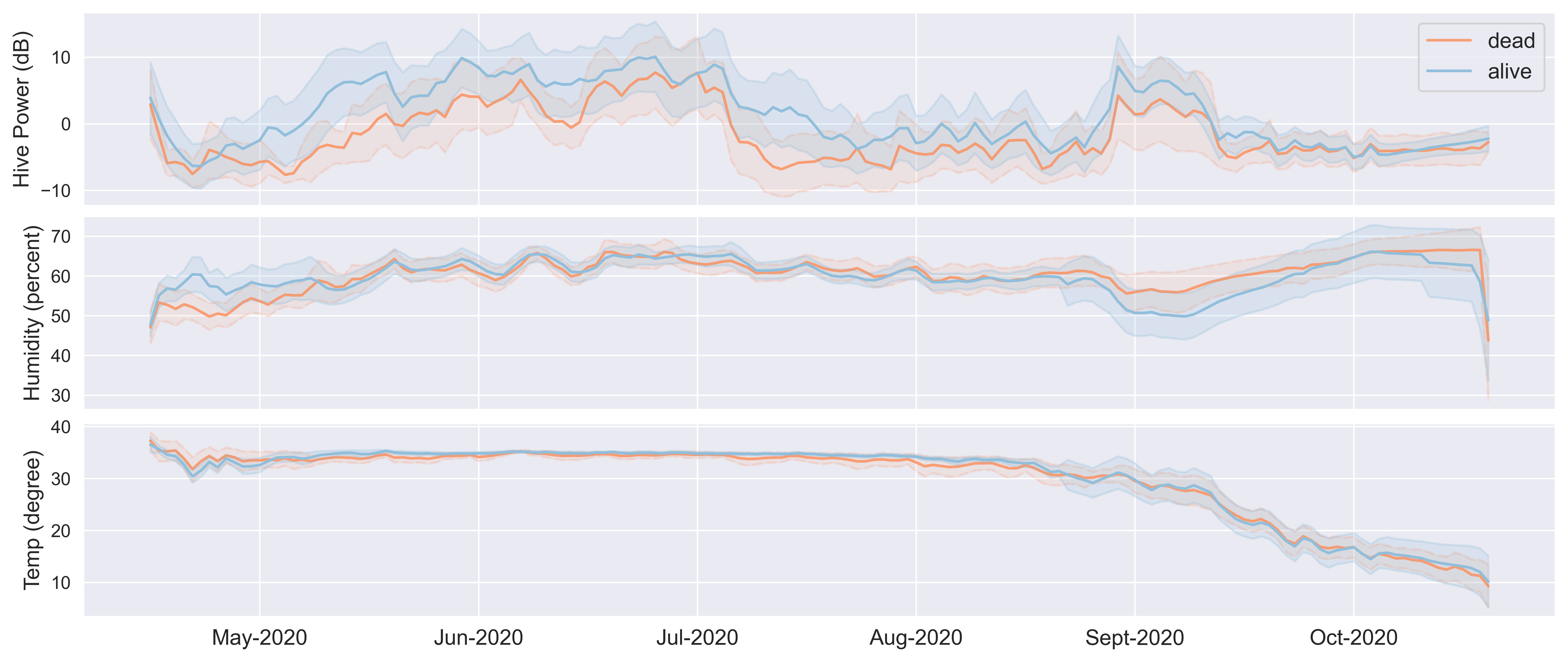}
\caption{Difference between hives that survived and failed in winter 2020-2021 based on multi-sensor data from May-2020 to October-2020. Data collected during wintering are not included in the plot. A statistical significant difference was found between the average hive power of failed and alive hives, while no significance was found for the other two modalities.}
\label{fig:diff_yearly}
\end{figure}

For the past fifteen years, winter loss of honeybee colonies have been observed throughout the world. In Canada, about 30\% of hives have failed during the winter months since 2007~\cite{capa, gray2020honey}. However, very limited tools can be used by beekeepers to identify the high-risk colonies at an early stage \cite{capa, gray2020honey}. Our exploratory analysis, in turn, has shown that hives that failed during the winter manifested a different pattern in multi-modal sensor data compared to those that survived. Figure~\ref{fig:diff_yearly} depicts such change in behavior for the three signal modalities. As can be seen, the main difference is observed with the audio modality, where failed hives showed significantly lower average hive power than the survivors based on Welch's t-test (p-value $=0.028$), indicating that the failed hives were less active before entering the wintering room. With the relative humidity and temperature, however, no significant difference has been found (p-value $=0.824$ and $0.702$ respectively).

\begin{figure}
\centering
\includegraphics[width=\linewidth]{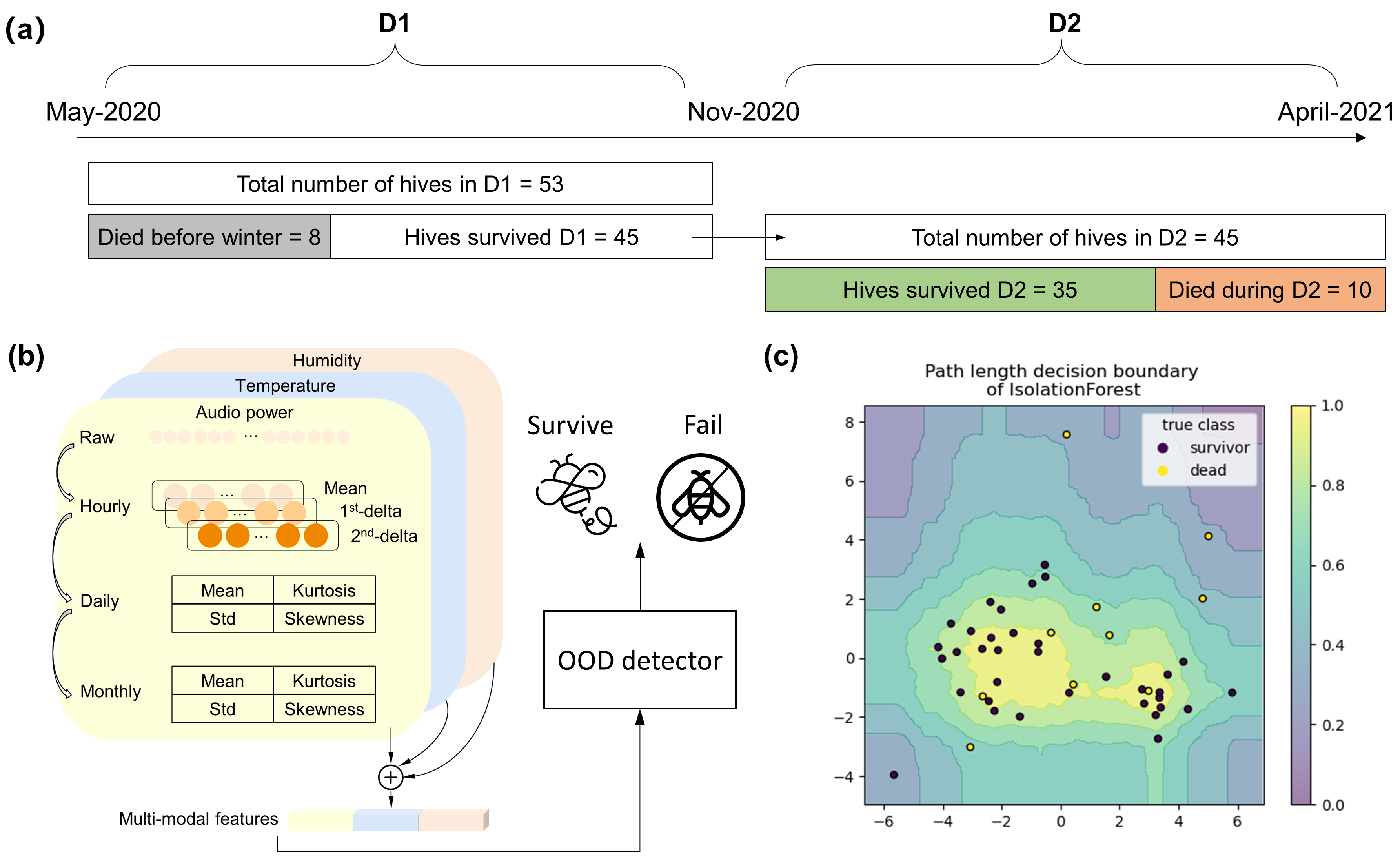}
\caption{Early-detection of beehive winter survival by using out-of-distribution detection models: (a) data partition and labelling, (b) feature extraction from multi-modal sensor data and downstream detection, and (c) decision boundary obtained from an isolation forest detector, where majority of the failed hives are found deviated from the distribution center, though a few overlapped with the alive hives.}
\label{fig:surv}
\end{figure}

In a recent study~\cite{zhu2023m}, the multi-modal sensor data (audio, relative humidity, temperature) was used to predict the beehive winter survivability using only features computed from the summer months.  In the experiments, all hives were divided into two categories: ones that survived the 2020-21 winter and ones that failed (see Figure~\ref{fig:surv}a). A hand-crafted feature was then proposed and derived from the multi-modal data, which aggregated the descriptors from different time-levels (Figure~\ref{fig:surv}b). For classification, parameter-free unsupervised out-of-distribution detection models were used (e.g., isolation forest with predefined parameters) to detect the hives that were less likely to survive the winter. The top-performing model achieved an AUC-ROC score of 0.730 based only on the data from May 2020 to October 2020. A visualization of the decision boundary obtained from the isolation forest classifier can be found in Figure~\ref{fig:surv}c. Overall, these findings suggest that multi-modal features can indeed encompass indicators of the hive winter survivability.

\subsection*{Population estimation}
\begin{figure}
    \centering
    \includegraphics[width=\linewidth]{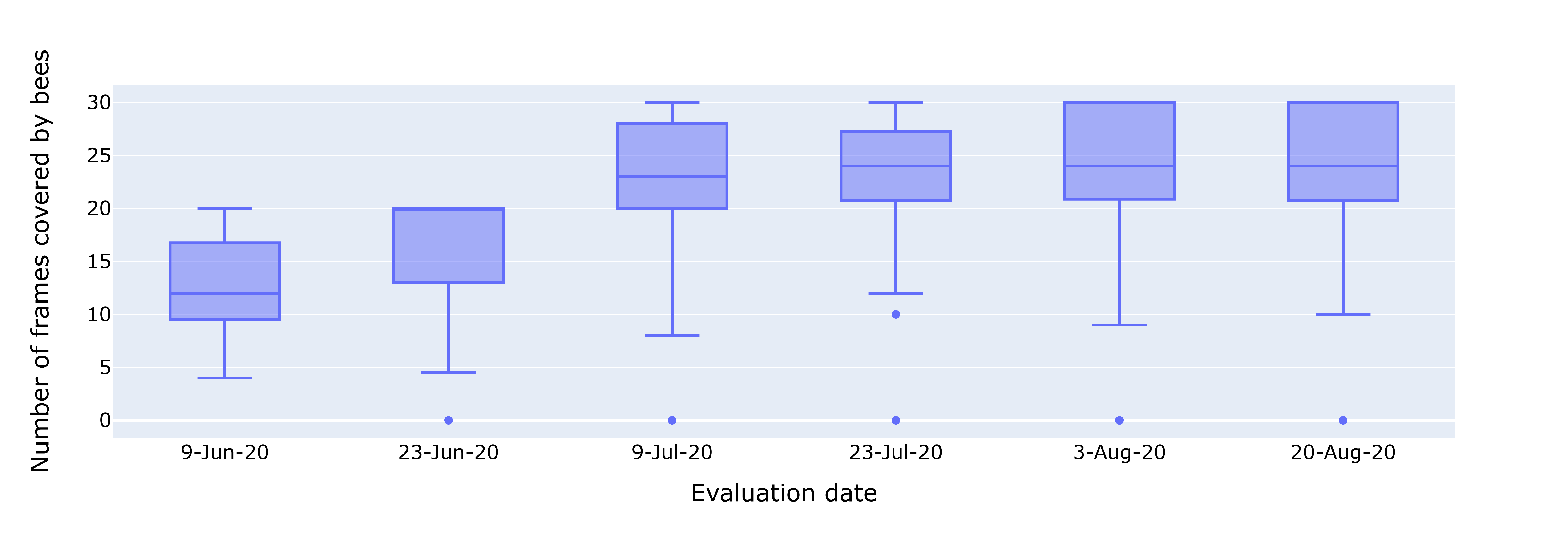}
    \caption{Distribution of the colony population from June, 2020 to August, 2020.}
    \label{fig:pop}
\end{figure}

As indicated in Figure~\ref{fig:yearly}, honey supers were added to each hive from June to August as colony populations increase and stock up their honey. The changes in hive population were quantified by counting the number of frames covered by bees (FoB) on a bi-weekly basis (see Figure~\ref{fig:pop}). The average FoB increases from 10 to 20 from June 9th to July 9th, then stabilizes at between 20 and 25 with the peak seen at the end of July. Meanwhile, it was observed that the population size varies markedly across hives. For example, while the majority (1st to 3rd quartile) had 20 to 30 FoB during July and August, as few as 10 FoB were seen in smaller colonies.

\begin{figure}
    \centering
    \includegraphics[width=0.4\linewidth]{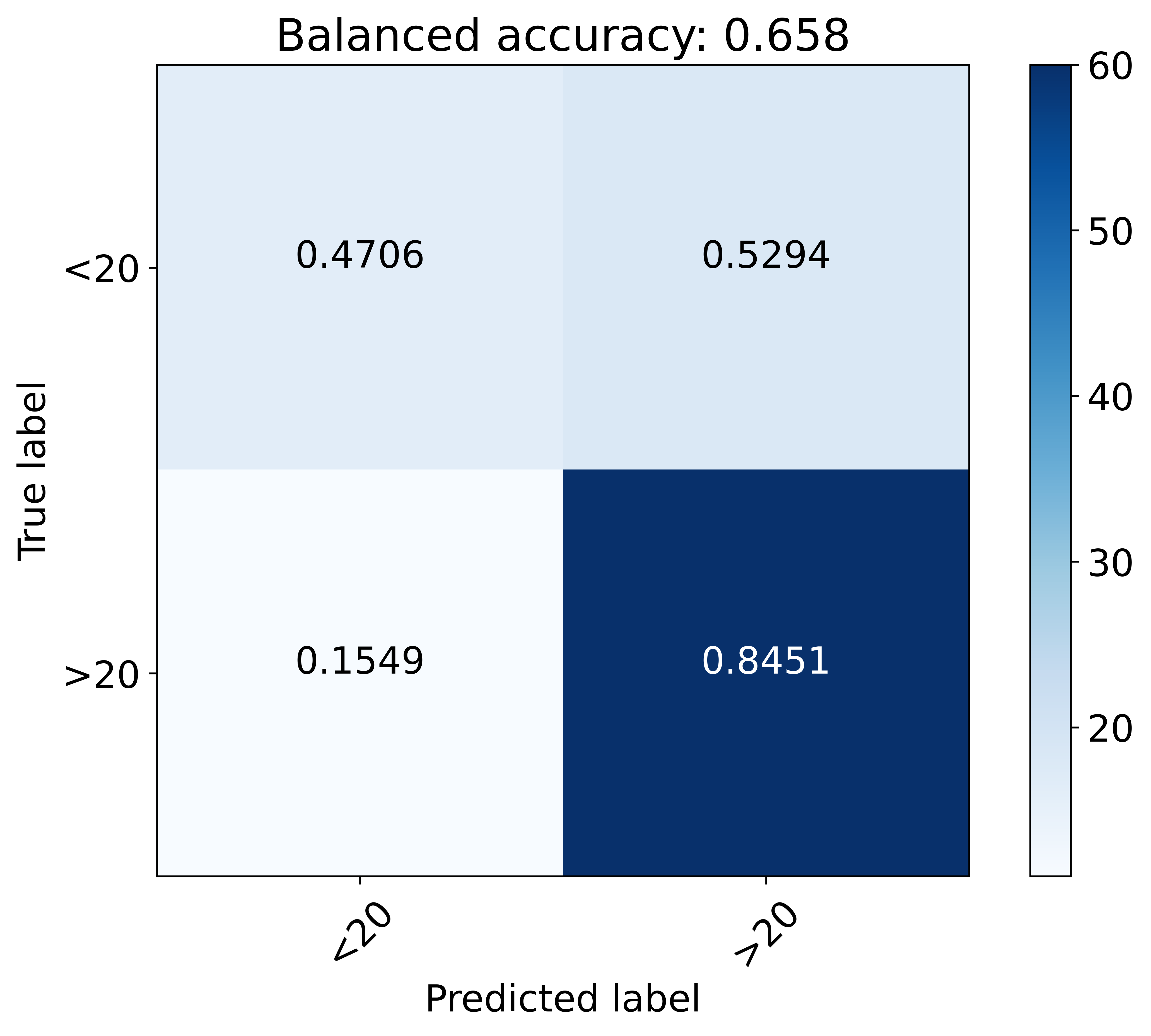}
    \caption{Confusion matrix for colony population size classification. Hives with over 20 FoB and below 20 FoB were separated into two classes. A linear SVM classifier obtained a balanced accuracy of 65.8\%.}
    \label{fig:cf}
\end{figure}

Recent studies have shown the usefulness of raw audio to predict hive population~\cite{abdollahi2022importance}. Here, we performed a similar task using features extracted from multi-modal sensor data to classify different population sizes. Colonies from the six evaluation dates were treated as independent samples, which resulted in a total number of 318 data points ($53 \times 6$). These were then divided into two classes based on the FoB, namely smaller hives comprised of FoB $<20$ and larger hives with FoB $>20$. We then extracted features (i.e., mean, 1st-order and 2nd-order deltas) from the sensor data three days before and after each evaluation date, and aggregated the daily features by computing low-level descriptors across the seven days (i.e., mean, standard deviation, kurtosis, and skewness). In the end, 24 features were computed per sample, which encapsulate information about the hive status close to the evaluation dates. These features were then fed into a linear Support Vector Machine (SVM) for classification. We followed a random data partition with a train-test-ratio of 7:3 and used 3-fold cross-validation to obtain the optimal classifier. Since the majority of the samples had FoB $>20$, we over-sampled the minority class to achieve a 1:1 ratio during training, and employed the balanced accuracy as the evaluation metric. As a result, the best classifier achieved a balanced accuracy of 65.8\%. Figure~\ref{fig:cf} reports the test results in a confusion matrix. Likely affected by the class imbalance, the classifier was found better at recognizing larger colonies (True Positive Rate = 84.5\%) and performed relatively worse with smaller colonies (True Negative Rate = 47.1\%).

\section*{Usage notes}

\subsection*{File reading and post-processing}
All data were shared as the comma separated values (CSV) files. The file reading can be easily done with Python using the Pandas library. The Python scripts used for post-processing, feature extraction, and machine learning tasks demonstrated in this paper can be found at our Github repository, listed below. 

\subsection*{Potential new use cases}
The present paper showcased the use of multimodal sensor data for a few different hive monitoring tasks. The data, however, is much richer and there are several new use cases that can lead to new insights about honeybee behavior. For example, while multiple modalities were collected, their fusion and the importance of each modality has yet to be quantified. Moreover, the phenotypic labels of the dataset can allow for many new ML applications to be developed, including but not limited to detection of \textit{Varroa} infestation, automated characterization of hygienic and defensive behaviors, or honey production prediction.

\section*{Code and media availability}
The code used for technical validation is available at our Github repository (\url{https://github.com/MuSAELab/MSPB}). Photos and videos of hive management are provided at the project webpage (\url{https://zhu00121.github.io/MSPB-webpage/}).

\bibliography{sample}


\section*{Acknowledgements} 
The authors acknowledge funding from NSERC via their Alliance program (ALLRP 548872-19), as well as Nectar Technologies Inc and the Centre de recherche en sciences animales de Deschambault for the support with data collection. 

\section*{Author contributions statement}
Y.Z. contributed to the initial drafting of the manuscript and conducted the experiments in the technical validation. M.A. contributed to data preparation and the initial drafting of the background section. S.M. contributed to phenotypic trait measurements collection, acquisition, and interpretation. N.C. contributed to sensor data acquisition and verified data records. H.G. contributed to the feedback to experimental results. P.G. and T.F. contributed in conceptualizing the study and study design. All authors contributed to the critical revision of the manuscript. All authors had full access to all the data in the study and took responsibility for the decision to submit this draft for publication.

\section*{Competing interests}
The authors declare no competing interests.


\end{document}